\input{aipcheck}

\def\apj{ApJ}

\def\mnras{MNRAS}
\def\pasp{PASP}

\def\apjs{ApJS}

\documentclass[
    ,final            
  ]
  {aipproc}

\layoutstyle{6x9}

\begin{document}

\title{Neutral Hydrogen and Star Formation in the Coma-Abell1367 Supercluster.}


\author{Luca Cortese}{
  address={School of Physics \& Astronomy - Cardiff University, The Parade, CF24 3AA Cardiff (UK)}
}

\begin{abstract}
We present preliminary results  of a multi-wavelength study focused on the evolution 
of spiral galaxies in the UV-optical colour-magnitude (CM) diagram. By combining HI, UV and optical observations 
of the Coma-Abell1367 supercluster we are able to identify galaxies at different stages of their evolution: from healthy star-forming galaxies, to 
blue HI-poor spirals and transition objects. Our analysis shows that galaxies in the transition region are likely 
to be the progeny of healthy spirals, whose star-formation has been quenched by the harsh cluster environment.
This result suggests that, at least in clusters of galaxies, the migration of galaxies from the blue to the red sequence 
might be due to environmental processes. 
\end{abstract}

\maketitle


\section{Introduction}

Star formation is probably the most fundamental of all astrophysical processes in galaxies. 
The overall evolution of galaxies depends on the rate at which their interstellar gas is converted into stars and the star formation rate depends on the rate at which diffuse interstellar matter is collected into star forming regions. 
Observations at high redshifts indicate that from $z\sim$1 to $z\sim$0 a large fraction of blue galaxies must have 
their star formation suppressed by some physical process, migrating from the blue to the red sequence (e.g. \cite{bell07}).
Different mechanisms have been invoked to explain this rapid quenching of the star formation activity, but so far none 
has been proven to satisfactorily reproduce the observations.
A possible step forward is represented by the combination of 
multi-wavelegth data able to trace the different ISM components that are involved in the star formation cycle: gas,  stars and dust.
This approach could allow us to put constraints on the evolutionary history of galaxies since different quenching mechanisms are supposed to influence the ISM components in different ways over different time-scales.

Here we report preliminary results  of a multi-wavelength analysis of the Coma-Abell1367 supercluster, focused on the 
study of the migration of spiral galaxies from the blue to the red sequence.

\section{The Coma-Abell1367 supercluster}
Abell1367 is an excellent place to study the impact of the galaxy environment on the transition from gas-rich to stellar 
dominated systems because it is a dynamically young cluster at the intersection of two Great Wall filaments.
Moreover the Arecibo Galaxy Environment Survey (AGES  \cite{auld06}) has just completed a HI blind survey 
of $\sim$5 deg$^{2}$ ($\sim$13 Mpc$^{2}$, assuming a distance of 92.8 Mpc) centered on the cluster core, providing crucial information on the HI content of cluster galaxies \cite{ages1367}.
The HI observations have been combined with UV (GALEX), optical (SDSS) and H$\alpha$ photometry publicly available for 
great part of the area surveyed by AGES.
\begin{figure}
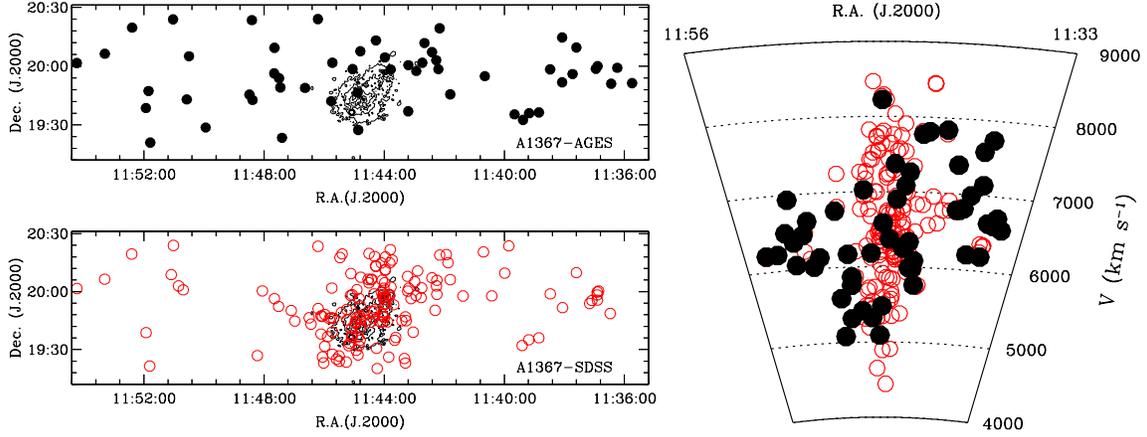

\centering
  \includegraphics[height=0.26\textheight]{cortese_fig1a.epsi}
  \includegraphics[height=0.26\textheight]{cortese_fig1b.epsi}
  \caption{The spatial distribution (left) and wedge diagram (right) for the HI (filled circles) and optically (empty circles) selected samples 
  in the Abell1367 supercluster (4000$<V<$9000 km s$^{-1}$).}
\end{figure}

\paragraph{HI-rich galaxies}
The sky distribution of the 54 HI sources detected by AGES (at a sensitivity limit of $\sim$6 10$^{8}$ M$_{\odot}$ for W$_{50}$=200 km s$^{-1}$) belonging to the Abell1367 region (4000$<V<9000$ km s$^{-1}$) is shown in Fig. 1 (left-hand panel, filled circles). No significant overdensity of sources is observed corresponding to the cluster centre. This is much more evident when we compare the sky distribution of our HI sample with that of optically selected galaxies obtained from SDSS-DR6 (empty circles). The optically selected sample is strongly clustered on the cluster centre and almost half of the detected sources lie within the X-ray emitting region. Moreover, the \emph{Finger of God} feature, typical of clusters of galaxies, is not observed in the HI selected sample (see Fig.1, right-hand panel). 
The differences observed between the optically and HI selected samples are somehow expected, since we usually associate 
HI galaxies with blue star-forming objects, quite rare in high density environments.
This is clearly shown in Fig.2 where we present the $FUV-i$ colour magnitude (CM) diagram (corrected for internal extinction following \cite{COdust05,afuv_luca}) for galaxies in our region: HI selected galaxies lie well within the blue sequence. 
We note that this is true only in a UV-optical colour magnitude relation, whereas in optical the blue and red sequences merge at high luminosities.

\paragraph{HI-poor galaxies}
Although all HI galaxies lie in the blue sequence, Fig.2 shows that not all blue sequence galaxies 
are HI-rich objects: $\sim$15-20\% of blue sequence galaxies are not detected in HI within our sensitivity limit.
These objects are all disks lying within a projected distance of $\sim$1 deg from the X-ray center of Abell1367.
Their broad-band morphology and structural parameters are indistinguishable from the other blue HI detected galaxies, whereas when 
observed in H$\alpha$ these objects show a truncation of the star-forming disc (see Fig.2), as usually observed in 
ram-pressure stripped galaxies \cite{review}.
These observational evidences suggest that we are probably dealing with objects that have just started their first dive into the core of Abell1367 and 
are gradually affected by the harsh cluster environment.

\section{Galaxy evolution in the CM diagram}
Can we speculate on the evolution of blue HI-poor galaxies?
A large fraction of their neutral hydrogen content has already been stripped, and the star formation has been 
quenched in the outskirts. Thus, it is likely that in less than a billion years these objects will become redder, migrating 
from the blue sequence and entering the transition region between the blue and red clouds.
This simple evolutionary scenario is supported by the properties of the transition galaxies in our sample.
They appear to be all disks, lying within $\sim$1 deg projected distance from the cluster center and showing 
a remarkable truncation in their star-forming disks (see Fig. 2). Transition galaxies are therefore likely to be  the progeny of the blue 
HI-poor objects, implying that the quenching of the star formation in disk galaxies could be related to environmental effects.
It has been recently shown that similar processes are probably behind the origin of the dwarf elliptical population of cluster galaxies 
\cite{dEale}, perhaps suggesting that the same physical mechanism is responsible for the migration of both massive and dwarf galaxies 
out of the blue sequence.

These results are not in contradiction with recent analysis showing that transitional galaxies preferentially host an active galactic nucleus (AGN, e.g. \cite{green}). 
It is in fact likely that our transition galaxies have AGN-like activity, although we do not have nuclear spectroscopy for all our sample. 
However, this is probably due to the fact that AGNs reside almost exclusively in high-mass systems \cite{agn,robyagn} and does 
not automatically imply that AGNs affect the star formation history of spiral galaxies. In fact, a relation between AGN feedback and 
quenching of the star formation in \emph{disk} galaxies still has to be proven \cite{okamoto08}.








\bibliographystyle{aipproc}   


\begin{figure}
  \includegraphics[height=0.9\textheight]{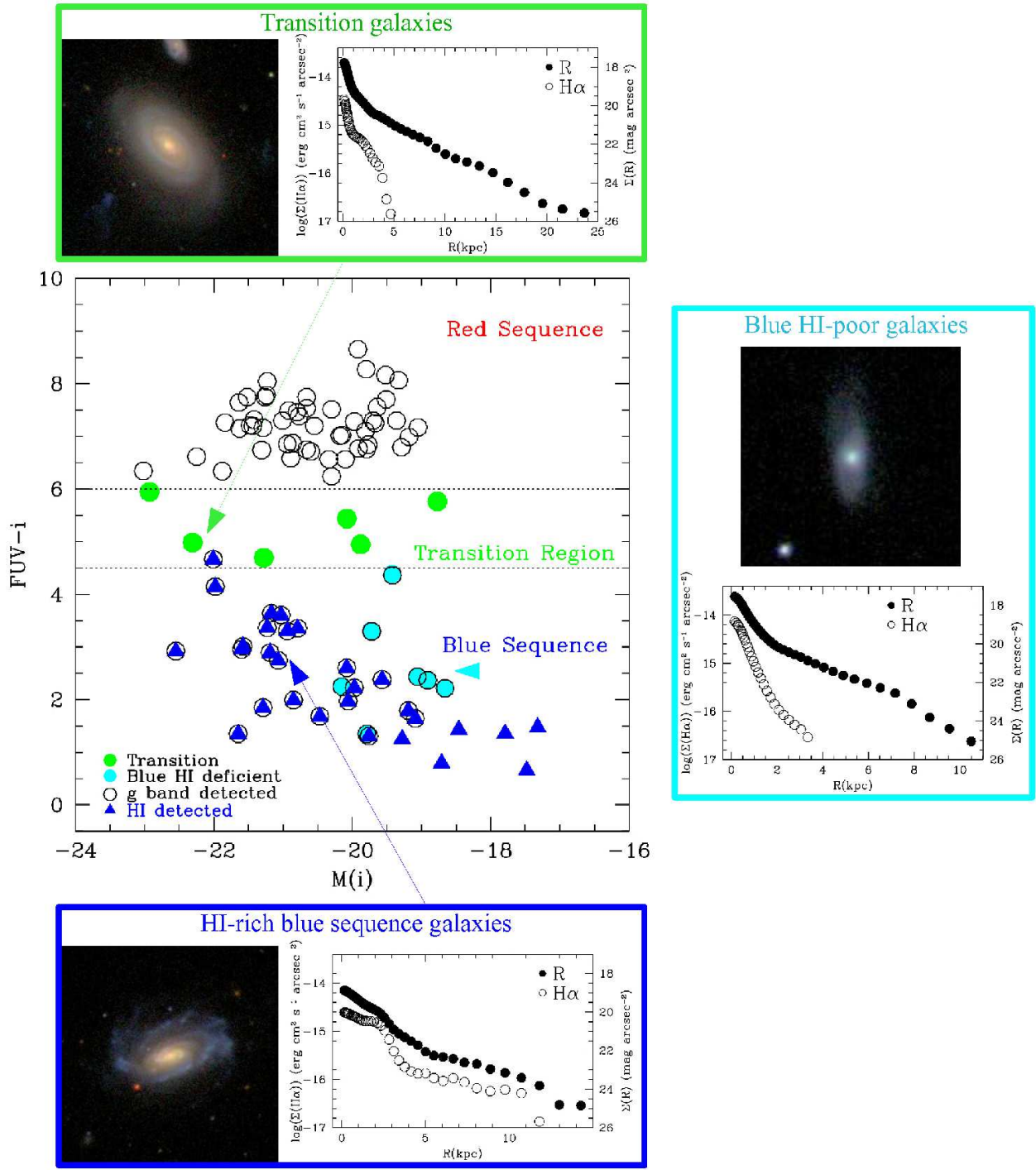}
  \caption{The $FUV-i$ colour magnitude relation for the AGES-Abell1367 field. Triangles and circles indicate HI and optically selected galaxies respectively.
   Blue HI-poor galaxies and transition systems are highlighted in cyan and green respectively.
  SDSS colour images, R band (filled circles) and H$\alpha$ (empty circles) surface brightness profiles are shown for a typical HI detected (lower panel), blue HI-poor (right-hand panel) and transition galaxy (upper panel).}
\end{figure}

\end{document}